\documentclass[floatfix, preprint, showpacs, showkeys, preprintnumbers, nofootinbib, superscriptaddress]{revtex4-1}
\usepackage{graphicx}% Include figure files
\usepackage{dcolumn}% Align table columns on decimal point
\usepackage{bm}% bold math
\usepackage{subfigure}
\usepackage{longtable}% bold math
\usepackage{setspace}
\usepackage{natbib}
\usepackage{amsmath}
\usepackage{rotating}
\usepackage{colortbl}
\usepackage{color}
\definecolor{mygray}{gray}{0.8}
\usepackage[colorlinks,citecolor=blue,linkcolor=blue,hypertex,breaklinks=true]{hyperref}

\begin{document}
%\title{Influence of neutron pairs oscillation around Fermi surface on the evolution of nuclear charge radii}
\title{Shell effects in nuclear charge radii based on Skyrme density functionals}
\author{Rong An}
%\email[Corresponding author: ]{rongan@nxu.edu.cn}
\affiliation{School of Physics, Ningxia University, Yinchuan 750021, China}
\affiliation{Guangxi Key Laboratory of Nuclear Physics and Technology, Guangxi Normal University, Guilin, 541004, China}
\affiliation{Key Laboratory of Beam Technology of Ministry of Education, School of Physics and Astronomy, Beijing Normal University, Beijing 100875, China}

\author{Shuai Sun}
\affiliation{Key Laboratory of Beam Technology of Ministry of Education, School of Physics and Astronomy, Beijing Normal University, Beijing 100875, China}

\author{Xiang Jiang}
\affiliation{College of Physics and Optoelectronic Engineering, Shenzhen University, Shenzhen 518060, China}

\author{Na Tang}
\affiliation{School of Physics, Ningxia University, Yinchuan 750021, China}
\affiliation{Key Laboratory of Beam Technology of Ministry of Education, School of Physics and Astronomy, Beijing Normal University, Beijing 100875, China}

\author{Li-Gang Cao}
\email[Corresponding author: ]{caolg@bnu.edu.cn}
\affiliation{Key Laboratory of Beam Technology of Ministry of Education, School of Physics and Astronomy, Beijing Normal University, Beijing 100875, China}
\affiliation{Institute of Radiation Technology, Beijing Academy of Science and Technology, Beijing 100875, China}

\author{Feng-Shou Zhang}
\email[Corresponding author: ]{fszhang@bnu.edu.cn}
\affiliation{Key Laboratory of Beam Technology of Ministry of Education, School of Physics and Astronomy, Beijing Normal University, Beijing 100875, China}
\affiliation{Institute of Radiation Technology, Beijing Academy of Science and Technology, Beijing 100875, China}
\affiliation{Center of Theoretical Nuclear Physics, National Laboratory of Heavy Ion Accelerator of Lanzhou, Lanzhou 730000, China}

\date{\today}

\begin{abstract}
 A unified description of the charge radii throughout the entire nuclide chart plays an essential role for our understanding of nuclear structure and fundamental nuclear interactions. In this work, the influence of new term, which catches the spirit of neutron and proton pairs condensation around Fermi surface, on the charge radii has been investigated based on the Skyrme density functionals with the effective forces SLy5 and SkM$^{*}$. The differential charge radii of even-even Ca, Ni, Sn, and Pb isotopes are employed to evaluate the validity of this theoretical model. Meanwhile, the results obtained by the relativistic density functional with the effective Lagrangian NL3 are also shown for the quantitative comparison. The calculated results suggest that the modified model can improve the trend of changes of the differential charge radii along Ca, Ni, Sn, and Pb isotopic chains, especially the shell closure effect at the neutron numbers $N=28$, 82 and 126. The shell quenching phenomena of charge radii can also be predicted at the neutron number $N=50$ along the corresponding Ni and Sn isotopes, respectively. The inverted parabolic-like shapes between the two fully filled shells can also be observed, but the amplitude is gradually weakened from Ca to Pb isotopic chains. Combining the existing literatures, it suggests that the discontinuous behavior in nuclear charge radii can be described well by considering the influence of neutron Cooper pairs condensation around Fermi surface.
\end{abstract}

%\pacs{25.70.Jj,24.10.-i}

\maketitle
\section{INTRODUCTION}\label{first}
Finite nuclei are self-consistently organized through two kinds of Fermion, namely the specific neutrons ($N$) and protons ($Z$) in a given nucleus.
As the strongly correlated many-body system, precise knowledge of nuclear charge radii gives raise to great attention in experimental and theoretical researches.
This is attributed to the fact that the regular and irregular variations in nuclear charge radii are intimately associated with our understanding of the fundamental interactions.
On one hand, under the assumption of the perfect charge symmetry, a highly linear correlation between the charge radii difference of mirror-paired nuclei and the slope parameter of symmetry energy has been built~\cite{PhysRevLett.132.162502,PhysRevLett.127.182503,PhysRevC.88.011301,PhysRevLett.119.122502,PhysRevResearch.2.022035}.
This seems to provide an alternative method in evaluating the equation of state (EoS) of isospin asymmetric nuclear matter.
On the other hand, as accurately measured quantity in terrestrial laboratories, charge radius plays an indispensable role in nuclear physics and astrophysics~\cite{ARNOULD2020103766,PhysRevC.108.015802}.
All of these motivations indicate that more charge radii data are urgently required in nuclear physics.

As encountered in nuclear masses, nuclear charge radii can also be directly detected in experiments.
Plenty of methods have been used to extract nuclear size, such as the high-energy elastic electron scattering ($e^{-}$)~\cite{PhysRev.92.978,PhysRevC.21.1426}, muonic atom x-rays ($u^{-}$)~\cite{ENGFER1974509,FRICKE1995177,BAZZI2011199}, optical isotopes shift (OIS)~\cite{PhysRevLett.65.1332,JBillowes_1995}, K$_{\alpha}$ x-rays isotopes shift (K$_{\alpha}$IS)~\cite{ANGELI2004185}, and highly-sensitive Collinear Resonance Ionization Spectroscopy (CRIS)~\cite{COCOLIOS2013565,Vernon2020}, etc.
However, charge radii of nuclei far away from the $\beta$-stability line cannot be easily measured owing to the unstable chemical properties. Over the past decade, laser spectroscopy method has been generally devoted to providing potential platform to extract the charge radii of exotic nuclei~\cite{CAMPBELL2016127,YANG2023104005}.
This is attributed to the underlying mechanism that the isotopic differences of the mean-square charge radii are
directly associated with the observable isotopic shifts of the hyperfine structure in atomic nuclei.
With the advanced techniques, much more data about charge radii of nuclei with larger neutron-to-proton ratios have been compiled~\cite{ANGELI201369,LI2021101440}.

Particularly, some local variations of nuclear charge radii can be generally observed, such as the shape-phase transition~\cite{PhysRevLett.117.172502,PhysRevC.95.044324,PhysRevC.99.044306,PhysRevLett.127.192501,PhysRevLett.128.152501,An2023035301}, shell quenching effects~\cite{PhysRevC.100.044310,BAGCHI2019251,PhysRevLett.129.142502,PhysRevC.104.064313,PhysRevC.102.051303,PhysRevLett.126.032502}, and the odd-even staggering (OES) phenomena~\cite{ANGELI201369,LI2021101440,PhysRevC.105.014325}.
Across the traditional neutron magic number $N=20$, as shown in Refs.~\cite{GarciaRuiz:2019cog,PhysRevC.105.L021303}, the rapid increase of the charge radii cannot be obviously performed along the calcium isotopic chain.
The same scenarios can also be encountered in the adjacent Ar and K isotopic chains ~\cite{KLEIN19961,TOUCHARD1982169,PhysRevLett.79.375,PhysRevC.92.014305}.
Recent study suggests that the strikingly abrupt increase of charge radius along the scandium isotopic chain has been observed across the neutron number $N=20$~\cite{PhysRevLett.131.102501}.
The pronounced discrepancies between the theoretical predictions and experimental databases can be encountered in describing the systematic evolution of the charge radii of nuclei far away from the $\beta$-stability line.
Thus more motivated underlying mechanisms should be captured properly in describing the finite size of atomic nuclei.

Various theoretical models are undertaken to characterize the range of nuclear charge radii, from the phenomenological formulae~\cite{PhysRevC.88.011301,Bohr1969,Zhang:2001nt,Sheng2015}, sophisticated mean-field models~\cite{geng2003,PhysRevLett.102.242501,PhysRevC.93.034337}, local-relation based approaches~\cite{PhysRevC.90.054318,PhysRevC.94.064315,PhysRevC.95.014307,PhysRevC.102.014306},
no-core shell model~\cite{PhysRevC.79.021303}, to the artificial neural network models~\cite{Utama_2016,PhysRevC.101.014304,PhysRevC.102.054323,PhysRevC.105.014308,DONG2023137726}.
The fine structures of nuclear charge radii can be influenced by various mechanisms~\cite{Angeli_2015}.
Particularly, the short-range correlations between neutrons and protons can have an influence on determining the proton density distributions and the occupation probabilities of single-particle levels around Fermi surface~\cite{PhysRevC.53.1599,MILLER2019360,PhysRevC.101.065202}.
This suggests that the neutron-proton correlations should be considered appropriately in describing the bulk properties of finite nuclei~\cite{COSYN2021136526,PhysRevC.107.064321}.
Moreover, the neutron-proton correlations derived from the associated valence neutrons and protons, namely being the Casten factor, can reproduce the shell quenching phenomena in nuclear charge radii as well~\cite{Sheng2015,PhysRevLett.58.658,Angeli_1991,Dieperink2009}.
Based on the relativistic mean-field model, a motivating method similar to the Casten factor has been proposed to capture the correlations between the neutrons and protons around Fermi surface, in which a novel ansatz derived from the neutron and proton pairs condensation has been incorporated into the root-mean-square (rms) charge radii formula~\cite{PhysRevC.102.024307,An_2022}.
To melt the overestimated OES phenomena in the charge radii, the modified version has been further proposed through considering the correlation between the simultaneously unpaired neutron and proton around Fermi surface~\cite{PhysRevC.109.064302}.
This modified approach can further reproduce the systematic evolution of charge radii along a long isotopic chain, especially the corresponding odd-even oscillation behaviors and the shell closure effects.
%{\color{red}Furthermore, it should be mentioned that four-particle correlations derived from a
%strong coupling between neutron- and proton-pairing correlations (being $\alpha$-clusters)~\cite{ZAWISCHA1985309} or the four-body force~\cite{ZAWISCHA1987299} can actually have an influence on determining the local variations of nuclear charge radii.}
With considering the density gradient terms in its surface pairing interactions part, the discontinuous behavior of charge radii can also be described well by the Fayans energy density functional (EDF)~\cite{PhysRevC.95.064328,PhysRevLett.122.192502}.

The bulk properties of finite nuclei can be described well at the mean field level~\cite{RevModPhys.75.121}.
However, as demonstrated in Refs.~\cite{SHARMA19939,PhysRevLett.128.022502,PhysRevC.110.054315}, the discontinuous variations of nuclear charge radii, such as the shell closure effects and OES phenomena, cannot be reproduced well by the Skyrme energy density functionals (EDFs).
It should be mentioned that such systematic description has become possible very recently owing to the development of a potential method to extract the discontinuous variations of nuclear charge radii in relativistic mean-field model~\cite{PhysRevC.102.024307,PhysRevC.109.064302}.
Therefore the extended application is further expected to provide a unified description of the charge radii under the non-relativistic Skyrme EDFs.
In this work, the Skyrme Hartree-Fock-Bogoliubov (HFB) framework within spherical symmetry~\cite{BENNACEUR200596} is employed and the differential mean-square charge radii of the even-even Ca, Ni, Sn, and Pb isotopes are calculated to quantify this modified model.
Furthermore, the results obtained by the relativistic EDFs are also shown for the quantitative comparison.

The structure of the paper is given as the following. In Section~2, the theoretical framework of Skyrme EDF and the modified rms charge radii formula are briefly presented. In Section~3, the numerical results and discussion are provided. Finally, a summary is given in Section~4.

\section{THEORETICAL FRAMEWORK}\label{second}
In the calculations, we use a Skyrme energy density functional (EDF), which has made considerable success to describe various phenomena in the course of nuclear physics~\cite{RevModPhys.75.121,PhysRevC.5.626,SAGAWA2001755,GRASSO2002103,PhysRevC.67.034313,PhysRevC.68.064302,PhysRevC.70.024307,PhysRevC.82.035804,PhysRevC.86.054313,PhysRevC.87.051302,PhysRevC.94.044313,PhysRevC.95.014316}.
The Skyrme effective interaction has been recalled as follows~\cite{CHABANAT1997710,CHABANAT1998231},
\begin{eqnarray}
V(\mathbf{r}_{1},\mathbf{r}_{2})&=&t_{0}(1+x_{0}\mathbf{P}_{\sigma})\delta(\mathbf{r})\nonumber\\
&&+\frac{1}{2}t_{1}(1+x_{1}\mathbf{P}_{\sigma})\left[\mathbf{P}'^{2}\delta(\mathbf{r})+\delta(\mathbf{r})\mathbf{P}^{2}\right]\nonumber\\
&&+t_{2}(1+x_{2}\mathbf{P}_{\sigma})\mathbf{P}'\cdot\delta(\mathbf{r})\mathbf{P}\nonumber\\
&&+\frac{1}{6}t_{3}(1+x_{3}\mathbf{P}_{\sigma})[\rho(\mathbf{R})]^{\alpha}\delta(\mathbf{r})\nonumber\\
&&+\mathrm{i}W_{0}\mathbf{\sigma}\cdot\left[\mathbf{P}'\times\delta(\mathbf{r})\mathbf{P}\right].
\end{eqnarray}
Here, $\mathbf{r}=\mathbf{r}_{1}-\mathbf{r}_{2}$ and $\mathbf{R}=(\mathbf{r}_{1}+\mathbf{r}_{2})/2$ are naturally related to the positions of two nucleons $\mathbf{r}_{1}$ and $\mathbf{r}_{2}$, $\mathbf{P}=(\nabla_{1}-\nabla_{2})/2\mathrm{i}$ represents the relative momentum operator acting on the right and the corresponding counterpart $\mathbf{P'}$ represents its complex conjugate acting on the left, and $\mathbf{P_{\sigma}}=(1+\vec{\sigma}_{1}\cdot\vec{\sigma}_{2})/2$ is the spin exchange operator that be used to control the relative strength of the $S=0$ and $S=1$ channels for a given term in the two-body
interactions, with $\vec{\sigma}_{1(2)}$ being the Pauli matrices.
The last term represents the spin-orbit force where $\sigma=\vec{\sigma}_{1}+\vec{\sigma}_{2}$, $W_{0}$ denotes the strength of the spin-orbit force.
The quantities $\alpha$, $t_{i}$ and $x_{i}$ ($i=0$-3) represent the parameters of the effective Skyrme forces used in this work.

The pairing correlations can be generally treated either by the BCS method or by the Bogoliubov transformation~\cite{PhysRevLett.102.242501,PhysRevC.82.035804,RYSSENS2015175}.
In this work, the Bogoliubov transformation is used to treat the pairing correlations under the Skyrme EDFs.
The density-dependent zero-range pairing force is employed as follows~\cite{PhysRevC.53.2809,PhysRevC.60.064312,Dobaczewski2002,PhysRevC.73.044309,MATSUO2007307,PhysRevC.86.054313,PhysRevC.80.044328},
\begin{eqnarray}
V_{\mathrm{pair}}(\mathbf{r}_{1},\mathbf{r}_{2})=V_{0}\left[1-\eta\left(\frac{\rho(\mathbf{r})}{\rho_{0}}\right)\right]\delta(\mathbf{r}_{1}-\mathbf{r}_{2}).
\end{eqnarray}
Here, $\rho(\mathbf{r})$ is the baryon density distribution in coordinate space and $\rho_{0}=0.16~\mathrm{fm}^{-3}$ represents the nuclear saturation density. Generally, the values of $\eta$ are taken as $0.0$, $0.5$, or $1.0$ for volume-, mixed-, or surface-type pairing interactions, respectively.
The mixed-type pairing force is chosen in our calculations.
The quantity $V_{0}$ is adjusted by calibrating the empirical pairing gaps with three-point formula~\cite{PhysRevC.86.054313,Bender:2000xk}. The single-particle energy levels and wave functions of constituent nucleons can be obtained by solving the HFB equations with the self-consistent iteration method.

The quantity of nuclear charge radius ($R_{\mathrm{ch}}$) deduced from the wave functions of the constituent protons can be generally used to define the range of charge density distributions.
Generally, the proton density distributions depends on the specific nucleon-nucleon interaction, i.e., the corresponding effective forces derived from the bulk properties of finite nuclei or nuclear matter.
 In determining the charge radii, the modified expression can be used as follows (in units of fm$^{2}$)~\cite{PhysRevC.109.064302},
\begin{eqnarray}\label{cp1}
R_{\mathrm{ch}}^{2}=\langle{r_{\mathrm{p}}^{2}}\rangle+0.7056+\frac{a_{0}}{\sqrt{A}}\Delta\mathcal{D}.
\end{eqnarray}
The first term $\langle{r_{\mathrm{p}}^{2}}\rangle$ represents the charge density distributions of point-like protons and the second one is attributed to the finite size of protons~\cite{Gambhir:1989mp,RevModPhys.93.025010,PhysRevLett.128.052002}.
For the third term, the expression $\Delta\mathcal{D}$ is defined as $\Delta\mathcal{D}=|\mathcal{D}_{n}-\mathcal{D}_{p}|$. The quantity of $\mathcal{D}_{n}$ ($\mathcal{D}_{p}$) has been defined as follows,
\begin{eqnarray}\label{cp2}
\mathcal{D}_{n,p}=\sum_{k>0}^{n,p} u_k{v}_k,
\end{eqnarray}
where $v_k^{n,p}$ is the amplitude of the occupation probability of the $k$th quasi-particle orbital for neutron or proton at the canonical basis, and $v_k^2={1-u_k^2}$. It is also mentioned that the quantity of $\mathcal{D}_{n,p}$ can be used to measure the Cooper pairs condensation around Fermi surface~\cite{PhysRevC.76.011302,PhysRevC.103.L021601,broglia2022}.
The expression of $\Delta\mathcal{D}$ has been used to measure the difference of neutron-pairs and proton-pairs condensation around Fermi surface~\cite{PhysRevC.105.014325,PhysRevC.102.024307}.
The parameter set $a_{0}=0.561$ is adjusted by reproducing the inverted parabolic-like shape and odd-even oscillation behaviors in charge radii of potassium and calcium isotopes~\cite{PhysRevC.109.064302}.
In our calculations, the same parameter set $a_{0}$ shown in Eq.~(\ref{cp1}) is further applied to depict the local variations of nuclear charge radii within Skyrme EDFs.
The time-reversal symmetry is assumed in this work, namely restricting ourselves to even-even nuclei.

\section{RESULTS AND DISCUSSION}\label{third}
In this work, the neutron-pairs condensation correction around Fermi surface are firstly incorporated into the root-mean-square (rms) charge radii formula based on the Skyrme EDFs.
The differential mean-square charge radii of calcium, nickel, tin, and lead isotopes are used to inspect the validity of this modified approach. To facilitate the universality of our results, the mixed-type pairing interactions are employed and the corresponding pairing strengthes used in this work are adjusted by fitting the empirical pairing gaps of the corresponding nuclei $^{44}$Ca, $^{64}$Ni, $^{122}$Sn, and $^{198}$Pb as shown in Table~\ref{tab0}.
For the convenience of discussion, the theoretical results obtained by Eq.~(\ref{cp1}) are labeled by HFB$^{*}$. While the theoretical results obtained by the approach without considering the neutron-pairs condensation correction term are marked by HFB.

\begin{table}[htbp!]
\centering
\caption{Pairing strength $V_{0}$ for Ca, Ni, Sn, and Pb isotopic chains used in this work (in units of MeV fm$^{3}$).}\label{tab0}
%\doublerulesep 0.1pt
\tabcolsep 13pt
\begin{tabular}{lcccr}
\hline
\hline
Sets   &  Ca   &  Ni   &   Sn     &  Pb~ \\
   \hline
 SLy5 & 245.2     &  362.8   &  410.0  &  468.0  \\
 SkM$^{*}$& 262.8 &  313.5  & 366.2   & 396.0\\
\hline\hline
\end{tabular}
\end{table}
%\subsection{Shell closure effect in nuclear charge radii}
With the increasing neutron numbers, the isotopic dependence of charge radii is distorted by the fully filled neutron shells.
This leads to the striking kink phenomena in nuclear charge radii due to the rather small isospin dependence of spin-orbit interactions~\cite{PhysRevLett.74.3744,REINHARD1995467}.
Actually, it is interesting to note that shell closure effects are generally observed in charge radii throughout the whole nuclide chart~\cite{ANGELI201369,LI2021101440}.
However, as mentioned before, Skyrme density functional fails to reproduce the shell quenching effect in nuclear charge radii~\cite{SHARMA19939,PhysRevLett.128.022502}. Recently developed RMF(BCS)$^{*}$ model with considering the correlation between the neutron and proton-pairs condensation around Fermi surface can reproduce the discontinuous variations in nuclear charge radii, especially for fully filled shells and the odd-even staggering phenomena~\cite{PhysRevC.109.064302}.
Therefore it is worthwhile to extend this approach to the Skyrme density functionals. In this work, the two representative parameter sets of effective forces SLy5~\cite{CHABANAT1998231} and SkM$^{*}$~\cite{BARTEL198279} are used, respectively.
Generally, these two parameter sets are still employed to investigate various nuclear structure phenomena~\cite{PhysRevC.110.064318,PhysRevC.110.064312,PhysRevC.110.064317,PhysRevResearch.7.013050,PhysRevC.111.014313}.
Here, nuclei with odd-neutron numbers are excluded from our discussion owing to the presence of the odd-even staggering effects.
To make a quantitative comparison, the results obtained by the relativistic EDF are necessary.
The relativistic-Hartree-Bogoliubov (RHB) model has been devoted to characterizing the bulk properties of finite nuclei as well~\cite{Kucharek1991,MENG19983,PhysRevC.82.011301}.
Recent study suggests that the RHB model with considering the modified term in the charge radii formula can describe the shell closure effect well~\cite{PhysRevC.110.064314}.
In this work, the RHB model within spherical symmetry is taken into account and the effective Lagrangian NL3 is used~\cite{PhysRevC.55.540}. In relativistic EDF model, the theoretical results obtained by Eq.~(\ref{cp1}) are labeled by RHB$^{*}$. While the results obtained by the approach without considering the neutron-pairs condensation correction term in the nuclear charge radii formula are labeled by RHB.

The abnormal behavior in nuclear charge radii can be profoundly observed along calcium isotopes.
The well-known inverted parabolic-like shape of charge radii can be significantly exhibited between the $^{40}$Ca and $^{48}$Ca isotopes~\cite{Ruiz2016}. Moreover, anomalous increase of charge radii can also be depicted across the neutron number $N=28$.
Furthermore, charge radii of nickel isotopes have been detected accurately through the advanced collinear laser spectroscopy method~\cite{PhysRevLett.127.182503,PhysRevLett.128.022502,PhysRevLett.129.132501}. From these detected results, one can find that the shrunken trend in the charge radii can be observed significantly around the fully filled $N=28$ shell, namely being the kink phenomenon. On the other hand, across the neutron number $N=28$, the systematic trend of changes in the charge radii is similar to calcium isotopic chain. It is also worth noting that the evolution of nuclear charge radii across the neutron number $N=28$ shows the universal trend from potassium to zinc isotopic chains, namely that is almost independent of the atomic number~\cite{PhysRevC.109.064302,PhysRevC.105.L021303}.

\begin{figure}[htbp!]
\centering
\includegraphics[width=1.0\linewidth]{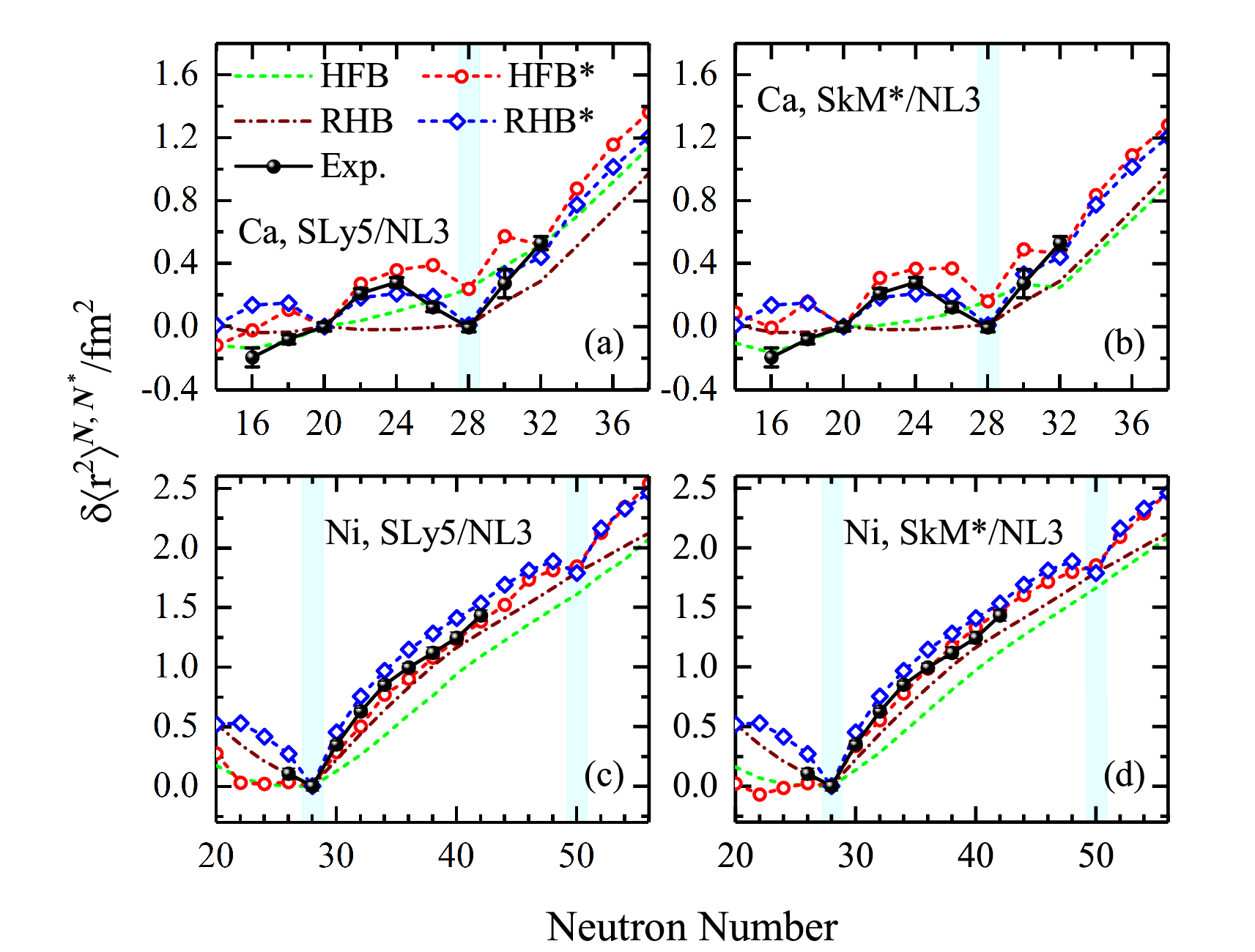}
   \caption{(Color online) Mean-square charge radii differences of the even-even Ca and Ni isotopes relative to the references $^{40}$Ca and $^{56}$Ni nuclei in the Skyrme EDFs with the parameter sets (a, c) SLy5 and (b, d) SkM$^{*}$ are presented for HFB (dashed line) and HFB$^{*}$ (open diamond) methods. The results obtained by the RHB (dot-dashed line) and RHB$^{*}$ (open diamond) approaches with effective Lagrangian NL3 are also shown for the quantitative comparison. The experimental data are taken from the Refs.~\cite{PhysRevLett.127.182503,ANGELI201369,LI2021101440,PhysRevLett.128.022502,PhysRevLett.129.132501} (solid circle). The shadowed planes mark the neutron magic numbers $N=28$ and $N=50$ (light blue bands), respectively. } \label{fig1}
\end{figure}
Attempts to describe these local variations of charge radii and verify the validity of this modified approach, as shown in Fig.~\ref{fig1}, the differential charge radii of calcium and nickel isotopes with respect to reference nuclei $^{40}$Ca and $^{58}$Ni are depicted by the HFB and HFB$^{*}$ methods, respectively.
In addition, the theoretical results obtained by the RHB and RHB$^{*}$ approaches are also depicted in this figure, respectively.
It is obvious from the Fig.~\ref{fig1}~(a) that the differential charge radii of calcium isotopes obtained by the HFB model with effective force SLy5 are smoothly increased with the increasing neutron numbers. This means that the inverted parabolic shape of charge radii between the neutron numbers $N=20$ and $28$ cannot be reproduced well. Meanwhile, HFB model fails to reproduce the shell quenching phenomenon of charge radii across the neutron number $N=28$.
The results obtained by the HFB$^{*}$ model are in agreement with the experimental data well in the $^{40-44}$Ca isotopes.
The inverted parabolic-like shape of charge radii between the neutron numbers $N=20$ and $28$ can also be significantly observed through the HFB$^{*}$ model, but the slight deviations can be found for the values of $^{46,48}$Ca isotopes.
Across the neutron number $N=28$, the abrupt increase in the charge radii can be described by the HFB$^{*}$ model as well.
Although the characteristic changes of slope around $N=28$ can be reproduced by the HFB$^{*}$ model, the systematic deviations cannot be omitted owing to the overestimated charge radius of $^{46,48,50}$Ca isotope.
By contrast, the relativistic mean field model can reproduce this local variations quantitatively at the fully filled $N=28$ shell~\cite{PhysRevC.102.024307,PhysRevC.109.064302}.
The Fayans EDF based on considering the density gradient terms in its pairing part can reproduce these irregular behaviors of charge radii along the calcium isotopes~\cite{PhysRevC.95.064328}.
By combining the calculated data with the Refs~\cite{PhysRevC.102.024307,PhysRevC.109.064302}, as well as the data from Refs.~\cite{PhysRevC.95.064328,PhysRevC.110.054315}, one should be mentioned that the overestimated charge radii can be obtained toward neutron-deficient regions as well.

As shown in Ref.~\cite{CAURIER2001240}, the abnormal changes of charge radii in the calcium isotopes can be described well through the shell model calculations, which takes effectively into account the phonon excitation.
In Fig.~\ref{fig1}~(b), the results obtained by the effective force SkM$^{*}$ are also employed for comparison study.
The differential charge radii of calcium isotopes calculated by the HFB model with SkM$^{*}$ force are also shown the smoothly increasing trend.
Although the inverted parabolic shape of charge radii between $^{40}$Ca and $^{48}$Ca and the shell closure effect at the $N=28$ can be given by the HFB$^{*}$ model, the values of charge radii along $^{46-50}$Ca isotopes are systematically overestimated against the experimental data, namely the local behavior around $N=28$ is similar to those obtained by the SLy5 set.
Moreover, charge radii of $^{36,38}$Ca isotopes are also overestimated with the effective force SkM$^{*}$.

From this figure, one can find that the RHB model cannot reproduce the inverted parabolic-like shape of charge radii between the fully filled $N=20$ and $N=28$ shells along calcium isotopes. Meanwhile, as encountered in the HFB model, the rapidly increasing trend of charge radii cannot be described well across the neutron number $N=28$. The RHB$^{*}$ method can reproduce the inverted parabolic-like shape of charge radii between the neutron numbers $N=20$ and $28$ as well as the shell closure effect around $N=28$. However, toward neutron-deficient regions, the results obtained by the RHB$^{*}$ method are overestimated as same as the cases obtained by the HFB$^{*}$ model with effective forces SLy5 and SkM$^{*}$. The HFB$^{*}$ and RHB$^{*}$ methods represent the similar trend of changes of the differential charge radii across the neutron number $N=32$, but the slight deviation can be found when the effective force SLy5 is used in the HFB$^{*}$ approach. It should be mentioned that the distinguished deviations can be exhibited between the results obtained by the HFB$^{*}$ and RHB$^{*}$ methods around the neutron number $N=28$. This can be understood easily from the result where relativistic EDF with effective force NL3 can almost reproduce the charge radius of $^{48}$Ca~\cite{PhysRevC.109.064302,PhysRevC.110.064314}. By contrast, HFB$^{*}$ model gives the slightly overestimated charge radius for $^{48}$Ca with respect to the experimental one.

In Figs.~\ref{fig1}~(c) and (d), it is found that HFB$^{*}$ model with both effective forces SLy5 and SkM$^{*}$ can reproduce the abrupt trend of changes of charge radii across the $N=28$ shell closure in the nickel isotopes.
Here, one can note that the values of charge radii for $^{60,62}$Ni isotopes are slightly underestimated.
In contrast to HFB$^{*}$ model, the differential charge radii in nickel isotopes obtained by HFB model fail to reproduce the experimental ones. Especially, the rapid increase of charge radii cannot be featured well across the fully filled $N=28$ shell.
The same scenario can also be encountered around the neutron number $N=50$ where the HFB model cannot give the shrunken trend of changes in the charge radii. Conversely, shell quenching effect in the charge radii can be pronouncedly observed at the neutron number $N=50$ based on the HFB$^{*}$ model.
For the RHB method, the shrunken trend of changes of charge radii can be characterized at the neutron number $N=28$. The RHB$^{*}$ method can also reproduce the kink phenomenon of charge radii at the $N=28$ and the shell quenching phenomenon at the $N=50$ can be predicted as well along the nickel isotopic chain. However, this shrunken trend of changes of charge radii at the neutron number $N=50$ cannot be featured in the RHB model as well as in the HFB method. In contrast to HFB$^{*}$ method, the differential charge radii of even-even $^{58-70}$Ni isotopes obtained by the RHB$^{*}$ method are slightly overestimated in comparison to the experimental data. Beyond the neutron number $N=50$, both of RHB$^{*}$ and HFB$^{*}$ methods show the similar trend of changes in the differential charge radii of nickel isotopes.

As encountered in potassium and calcium isotopes, the parabolic-like shape of charge radii can be profoundly observed between the neutron numbers $N=20$ and $N=28$~\cite{ANGELI201369,LI2021101440}. In the HFB$^{*}$ and RHB$^{*}$ models calculations, the decreasing trend of charge radii from $^{54}$Ni to $^{56}$Ni can be described as well.
This leads to the evidently observed kink phenomenon at the neutron number $N=28$.
Combining the shrunken phenomenon of charge radii around the $N=50$, the inverted parabolic-like shape of charge radii can be found between the neutron numbers $N=28$ and $N=50$ along nickel isotopic chain.
The same scenario can also be found in the RHB$^{*}$ model.
However, these phenomena are vanished in the HFB and RHB models due to the shell closure effect of charge radii at the $N=50$ cannot be emerged from both of them.
By contrast to the inverted parabolic-like shape of charge radii in the potassium and calcium isotopes, the amplitude is apparently weakened between the neutron numbers $N=28$ and $N=50$ along Ni isotopic chain.
Besides, the distinguished deviations can be found between the results obtained by the relativistic and Skyrme EDFs toward neutron-deficient regions.
Thus more underlying mechanisms and experimental data are urgently required in the proceeding research.

\begin{figure}[htbp!]
\centering
\includegraphics[width=1.0\linewidth]{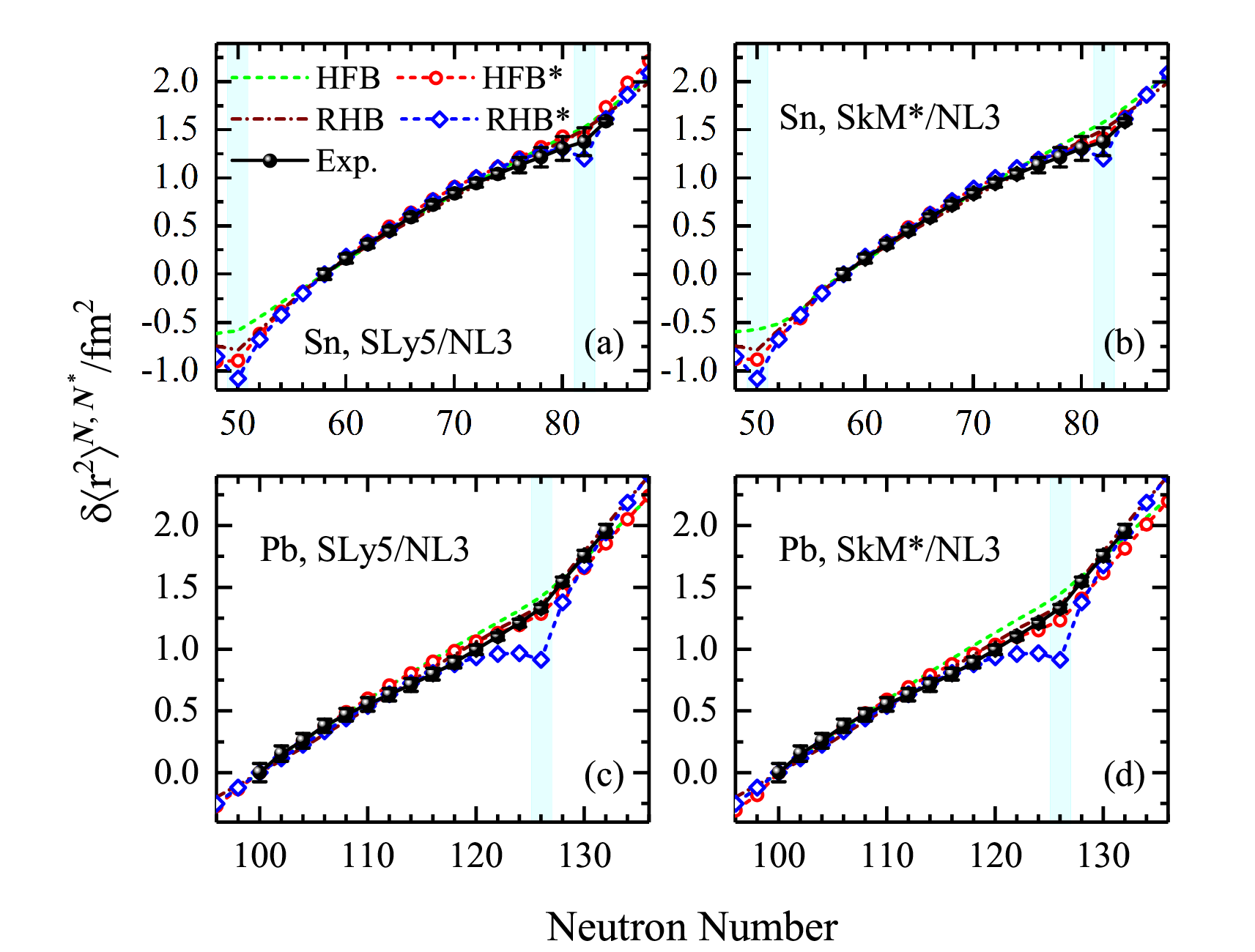}
   \caption{(Color online) Mean-square charge radii differences of the even-even Sn and Pb isotopes relative to the nuclei $^{108}$Sn and $^{182}$Pb in the Skyrme EDFs with the effective forces (a, c) SLy5 and (b, d) SkM$^{*}$ are presented for HFB (dashed line) and HFB$^{*}$ (open diamond) methods. The results obtained by the RHB (dot-dashed line) and RHB$^{*}$ (open diamond) approaches with effective Lagrangian NL3 are also shown for the quantitative comparison. The experimental data are taken from the Refs.~\cite{ANGELI201369,LI2021101440} (solid circle). The shadowed planes mark the neutron magic numbers $N=50$, $82$ and $N=126$ (light blue bands), respectively. } \label{fig2}
\end{figure}

To further examine the shell quenching phenomena of charge radii at the neutron numbers $N=82$ and $126$,
 as shown in Fig.~\ref{fig2}, the differential charge radii of the tin and lead isotopes are also depicted by the HFB and HFB$^{*}$ models with the effective forces SLy5 and SkM$^{*}$, respectively.
Meanwhile, the theoretical results obtained by the RHB and RHB$^{*}$ methods with effective force NL3 are also shown, respectively.
The HFB model gives the almost linear change in the slope around the traditional neutron number $N=82$ along tin isotopes.
The same scenario can also be encountered in the lead isotopes where the almost smooth trend of changes of charge radii can be obtained around the neutron number $N=126$.
These results suggest that the shrunken trend of charge radii (being marked kinks) at the fully filled $N=82$ and $N=126$ shells cannot be reproduced well through the HFB model.
By contrast, shell quenching phenomena of charge radii in the tin and lead isotopes can be described well by the HFB$^{*}$ model.
Here, one should be mentioned that the shell closure effect at the neutron number $N=82$ cannot be described well through the RHB model. In contrast to RHB model, the RHB$^{*}$ model gives an improved description in reproducing the kink phenomenon of charge radii at the nucleus $^{132}$Sn. However, the large deviations between the differential charge radii of even-even $^{202-206}$Pb isotopes obtained by the RHB$^{*}$ model and the experimental data can be significantly exhibited as shown in Fig.~\ref{fig2}. The shell effect of the differential charge radii at the nucleus $^{208}$Pb cannot be described well by the RHB$^{*}$ model.
As shown in Ref.~\cite{PhysRevC.102.024307}, the large deviations can be depicted between the charge radii of Pb isotopes calculated by the modified formula and the experimental data due to the overestimated parameter set $a_{0}$ shown in Eq.~(\ref{cp1}). In addition, the shape deformation may also have an influence on determining the charge radii of nuclei toward neutron-deficient region. More details can be found in the following discussion.
Moreover, shell closure effect at the $^{100}$Sn isotope can also be predicted by the HFB$^{*}$ and RHB$^{*}$ models.
This should be further confirmed due to the lack of experimental data in the neutron-deficient region.
Although the rapid increase of charge radii across $N=126$ can be reproduced by the HFB$^{*}$ model along the lead isotopes, the slope of changes are slightly deviated from the experimental data. Beyond the neutron numbers $N=82$ and $N=126$, the HFB$^{*}$ and RHB$^{*}$ models give the similar trends, namely the rapid increases of the differential charge radii along the tin and lead isotopic chains.

\begin{figure*}[htbp!]
\centering
\includegraphics[width=1.05\linewidth]{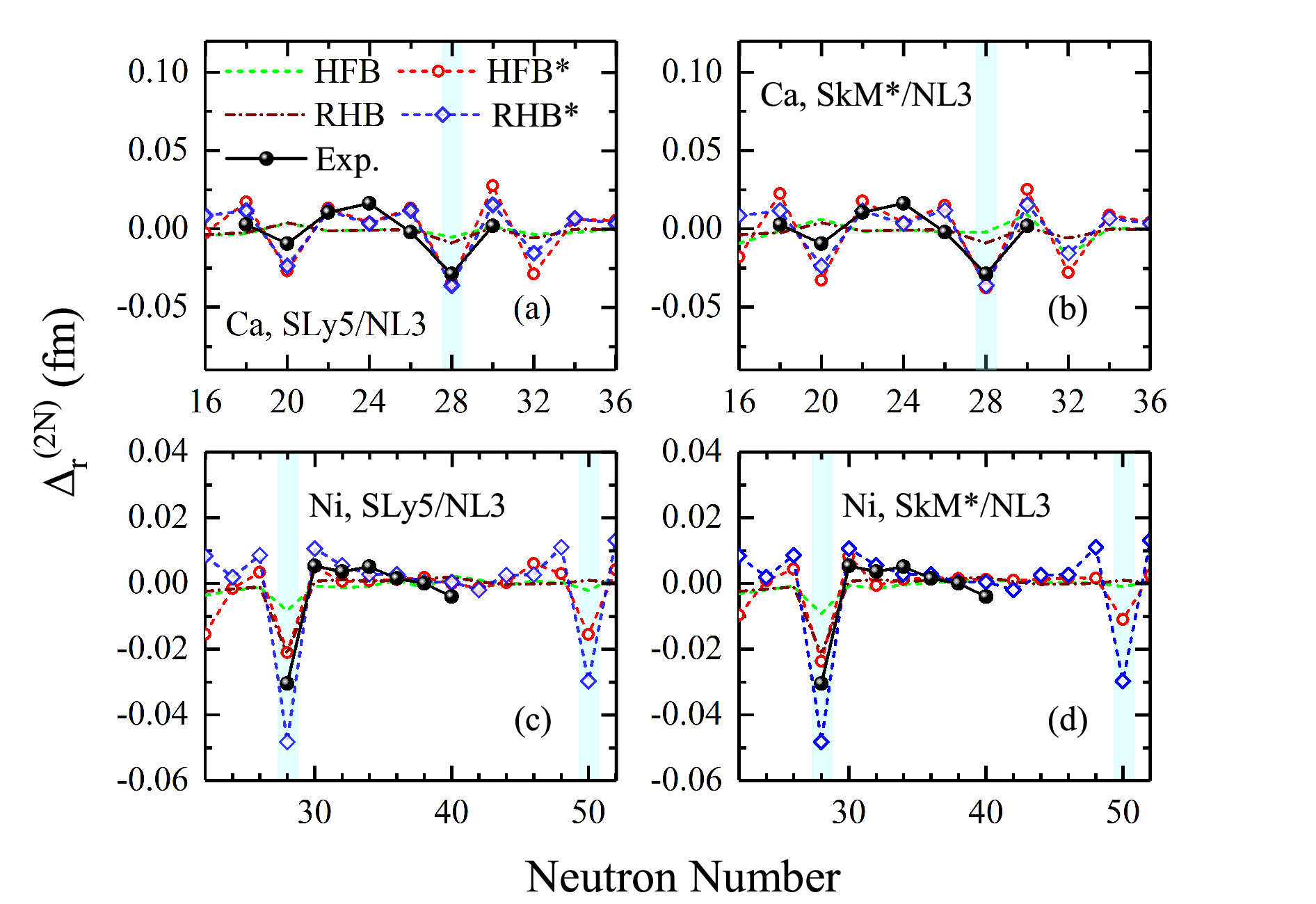}
   \caption{(Color online) The values of $\Delta_{r}^{(2N)}$ obtained by the HFB (dashed line) and HFB$^{*}$ (open diamond) methods are presented for charge radii of even-even (a, b) Ca and (c, d) Ni isotopes with the parameter sets SLy5 and SkM$^{*}$, respectively. The results obtained by the RHB (dot-dashed line) and RHB$^{*}$ (open diamond) approaches with effective Lagrangian NL3 are also shown for the quantitative comparison. The experimental data are taken from the Refs.~\cite{PhysRevLett.127.182503,ANGELI201369,LI2021101440,PhysRevLett.128.022502,PhysRevLett.129.132501} (solid circle). The neutron magic numbers $N=28$ and $50$ are marked by the light blue bands.} \label{fig3}
\end{figure*}

In order to feature the local variations of nuclear charge radii along a long isotopic chain, namely to eliminate the smooth mass number dependence of the charge radii, the three-point formula has been recalled as follows~\cite{PhysRevC.102.024307,PhysRevC.109.064302},
\begin{small}
\begin{eqnarray}\label{oef1}
\Delta_{r}^{(2N)}(N,Z)=\frac{1}{2}[2R(N,Z)-R(N-2,Z)-R(N+2,Z)],
\end{eqnarray}
\end{small}
where $R(N,Z)$ is rms charge radius for a nucleus with neutron number $N$ and proton number $Z$.
In Fig.~\ref{fig3}, the local variations of $\Delta_{r}^{(2N)}$ obtained by the HFB, HFB$^{*}$, RHB, and RHB$^{*}$ models are depicted along the even-even Ca and Ni isotopes.
The HFB model with effective force SLy5 cannot describe the experimental data as shown in Fig.~\ref{fig3}~(a).
The values of $\Delta_{r}^{(2N)}$ can be reproduced well by the HFB$^{*}$ model, but the slight deviation can be found at the $^{44}$Ca isotope. This can be understood easily from Fig.~\ref{fig1}~(a) where the evaluated differential charge radii of $^{46,48}$Ca isotopes are apparently deviated from the experimental data.
The same scenarios can also be encountered in the results obtained by the SkM$^{*}$ force, namely the values of $\Delta_{r}^{(2N)}$ cannot be reproduced by the HFB model.
Meanwhile, the trend of changes of $\Delta_{r}^{(2N)}$ at the neutron number $N=24$ is not accord with the experimental one.
This results from the overestimated differential charge radii of $^{46,48}$Ca isotopes and this can also be found from Fig.~\ref{fig1}.
Moreover, it should be mentioned that the HFB and HFB$^{*}$ models with the effective SkM$^{*}$ force give the similar trend of $\Delta_{r}^{(2N)}$ at the neutron number $N=32$ as shown in Fig.~\ref{fig3}~(b).
Although the differential charge radii of neutron-deficient calcium isotopes are overestimated, here it is worthwhile to mention that the systematic trend of $\Delta_{r}^{(2N)}$ can be reproduced by the HFB$^{*}$ model. Besides, the HFB method even gives the slightly inverse trend at the $N=20$.

For the nickel isotopes, the HFB model with effective forces SLy5 and SkM$^{*}$ cannot describe well the local trends of changes of $\Delta_{r}^{(2N)}$ as shown in Fig.~\ref{fig3}~(c) and (d).
From the values of $\Delta_{r}^{(2N)}$, one can find the shell quenching phenomena in the charge radii can be profoundly observed with the HFB$^{*}$ model.
Especially, the shell closure effect can also be presented at the $N=50$.
The decreased trend of $\Delta_{r}^{(2N)}$ from $^{58}$Ni to $^{60}$Ni can be reproduced by the HFB$^{*}$ model with both the effective forces SLy5 and SkM$^{*}$. The similar trend from $^{66}$Ni to $^{68}$Ni can also be given by the HFB$^{*}$ model with the effective forces SLy5, but slight deviation is presented with respect to the SkM$^{*}$ set.

As shown in Fig.~\ref{fig3}, one can find that the local trends of changes of $\Delta_{r}^{(2N)}$ obtained by the RHB model are similar to the cases obtained by the HFB model along the calcium isotopes and cannot reproduce the experimental data. The almost similar trend in the values of $\Delta_{r}^{(2N)}$ can also be drawn through the HFB$^{*}$ and RHB$^{*}$ methods.
Along Ni isotopic chains, the HFB$^{*}$ and RHB$^{*}$ methods show the inverse trend of changes of $\Delta_{r}^{(2N)}$ at the neutron number $N=22$. Furthermore, the values of $\Delta_{r}^{(2N)}$ obtained by the RHB$^{*}$ method are more larger than the results obtained by the HFB$^{*}$ model at the nuclei $^{56}$Ni and $^{78}$Ni. Around the neutron number $N=50$, the slightly deviations can also be presented between the HFB$^{*}$ and RHB$^{*}$ models.

\begin{figure*}[htbp!]
\centering
\includegraphics[width=1.05\linewidth]{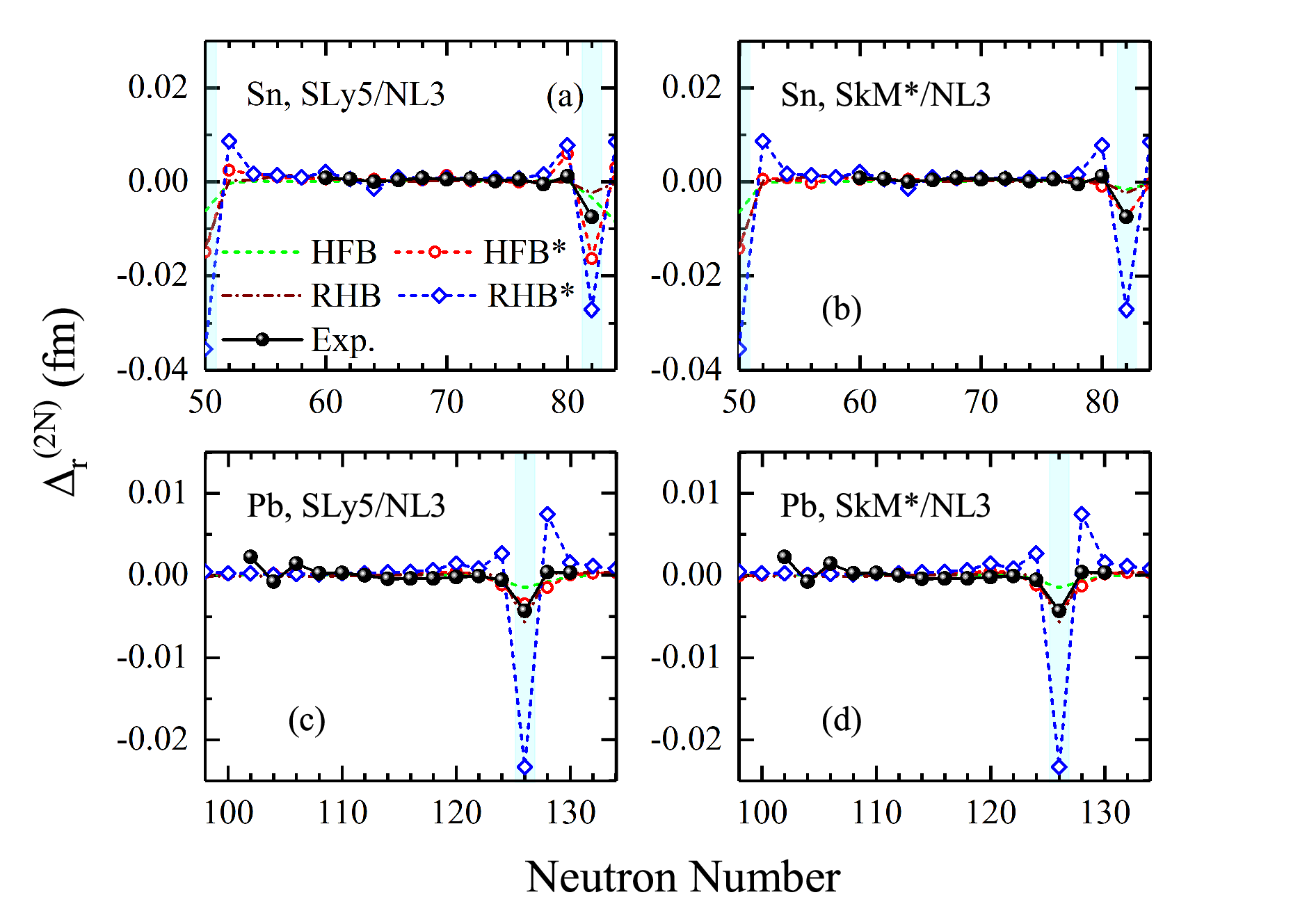}
   \caption{(Color online) The values of $\Delta_{r}^{(2N)}$ obtained by the HFB (dashed line) and HFB$^{*}$ (open diamond) methods are presented for charge radii of even-even (a, b) Sn and (c, d) Pb isotopes with the parameter sets SLy5 and SkM$^{*}$, respectively. The results obtained by the RHB (dot-dashed line) and RHB$^{*}$ (open diamond) approaches with effective Lagrangian NL3 are also shown for the quantitative comparison. The experimental data are taken from the Refs.~\cite{ANGELI201369,LI2021101440} (solid circle). The neutron magic numbers $N=50$, $82$ and $126$ are marked by the light blue bands.} \label{fig4}
\end{figure*}
To further review the discontinuous behavior in nuclear charge radii, the local variations of $\Delta_{r}^{(2N)}$ obtained by the HFB and HFB$^{*}$ models are depicted along the even-even Sn and Pb isotopes as shown in Fig.~\ref{fig4}.
The values of $\Delta_{r}^{(2N)}$ obtained by the RHB and RHB$^{*}$ approaches are also depicted, respectively.
Along the tin isotopes shown in Fig.~\ref{fig4}~(a) and (b), the HFB$^{*}$ and HFB models give almost similar trend of changes of $\Delta_{r}^{(2N)}$ except for the cases around the neutron numbers $N=50$ and $82$.
The HFB$^{*}$ model with the SLy5 parameter set can reproduce the trend of change of $\Delta_{r}^{(2N)}$ at the neutron number $N=82$, but the oscillated amplitude is overestimated.
It is important to note that the values of $\Delta_{r}^{(2N)}$ can be reproduced well by the HFB$^{*}$ model with effective force SkM$^{*}$ set.
This can also be found in Fig.~\ref{fig2} where the shrunken trend of changes of charge radii along Sn isotopes can be characterized well using the HFB$^{*}$ model with the SkM$^{*}$ set in comparison to the SLy5 set.
Moreover, the increased trend of $\Delta_{r}^{(2N)}$ from $^{128}$Sn to $^{130}$Sn can be obtained by the SLy5 parameter set.
Compared to SLy5 set, the HFB$^{*}$ model with SkM$^{*}$ set gives the slightly inverse trend in $\Delta_{r}^{(2N)}$ from $^{128}$Sn to $^{130}$Sn.
The RHB model gives the similar trend of changes of $\Delta_{r}^{(2N)}$ with respect to the results obtained by the HFB method, but the slight deviations can be shown for the nuclei $^{100}$Sn and $^{134}$Sn. In addition, the shell effect at the neutron number $N=50$ can also be featured through the value of $\Delta_{r}^{(2N)}$. This can be obviously seen from Fig.~\ref{fig2} where the kink phenomenon of the differential charge radii can be characterized well at the nucleus $^{100}$Sn. Besides, the values of $\Delta_{r}^{(2N)}$ obtained by the RHB$^{*}$ model are slightly overestimated around the nuclei $^{100}$Sn and $^{132}$Sn compared to the results calculated through HFB$^{*}$ model.

It is obvious from the Fig.~\ref{fig4}~(c) and (d) that the results obtained by the HFB$^{*}$ method with effective forces SLy5 and SkM$^{*}$ depict the shell closure effect well at the neutron number $N=126$.
Although the comparable $\Delta_{r}^{(2N)}$ values of $^{188-206}$Pb isotopes are obtained by both HFB and HFB$^{*}$ methods, HFB method fails to capture the shell quenching phenomenon at the $N=126$.
The results obtained by the RHB model show the similar trend of changes of $\Delta_{r}^{(2N)}$ with respect to calculations performed by the HFB model, but the value of $\Delta_{r}^{(2N)}$ at the neutron number $N=126$ can be reproduced well by the RHB model.
By contrast, the value of $\Delta_{r}^{(2N)}$ obtained by the RHB$^{*}$ model deviates from the experimental value of nucleus $^{208}$Pb. Meanwhile, the larger deviations are shown between the values of $\Delta_{r}^{(2N)}$ obtained by the RHB$^{*}$ and HFB$^{*}$ models around the neutron number $N=126$, respectively.
It is surprising that the slight oscillation behavior can also be presented in the values of $\Delta_{r}^{(2N)}$ toward neutron-deficient lead isotopes.
Both of the HFB$^{*}$ and HFB models cannot follow these discontinuous trends from the neutron numbers $N=102$ to $104$.
The same scenarios can also be encountered in the RHB and RHB$^{*}$ models.
As shown in Refs.~\cite{PhysRevC.95.044324,PhysRevC.99.044306,PhysRevLett.126.032502,PhysRevLett.127.192501,PhysRevC.104.024328,Nat1163}, around $Z = 82$ isotopic chains, the shape deformation plays an indispensable role in determining the local variations of nuclear charge radii toward neutron-deficient regions.
In our discussions, the deformation effect cannot be taken into account.
Although the HFB$^{*}$ model can make an improvement in describing the differential charge radii along calcium, nickel, tin, and lead isotopes, some local variations cannot be captured adequately yet as shown in Figs.~\ref{fig3} and \ref{fig4}.
Therefore more underlying mechanisms, such as the shape deformation, should be captured appropriately.

From the infinite nuclear matter to finite nuclei, the fundamental interactions offered by the effective forces cannot be captured adequately in atomic nuclei. This can be understood from some specific aspects where the radii of the proton density distributions can be significantly changed, such as shell closure effect~\cite{PhysRevC.104.064313,Bhuyan_2021}, shape deformation~\cite{PhysRevLett.126.032502,PhysRevLett.127.192501}, the influence coming from isospin symmetry breaking~\cite{PhysRevC.105.L021304,PhysRevC.106.L061306,PhysRevC.107.064302}, and so on.
For HFB$^{*}$ model, a greater ability to reproduce the differential charge radii of Ca, Ni, Sn, and Pb isotopes is attributed to the correction term derived from the neutron- and proton-pairs condensation between the states below and above the Fermi surface due to the superconductivity property.
Combining the existing results obtained by the relativistic mean-field model~\cite{PhysRevC.105.014325,PhysRevC.102.024307,PhysRevC.109.064302}, this means that the modification term derived from the neutron- and proton-pairs condensation highlights the importance of describing the local variations of nuclear charge radii well.
However, the incorporated term in Eq. (3) is actually phenomenological set, although the quantities $D_{n}$ and $D_{p}$ are obtained self-consistently from the microscopic aspect. This leads to the fact that the underlying mechanism is covered up by this parametrization set. The effect of neutron-proton short-range correlations on charge radii seems to be necessary to achieve a calculation of high precision~\cite{MILLER2019360}. This encourages us to investigate the charge radii of nuclei by performing HFB calculations with the pairing interaction in the $T = 0$ channel and eventually going beyond the mean field approach. 

As is well known, the neutron skin thickness is a superior indicator to ascertain the equation of state in isospin asymmetric nuclear matter.
The difference in proton density distributions of mirror nuclei is intimately associated to the neutron skin thickness~\cite{PhysRevC.97.014314,nuclscitech35.182}.
This provides an alternative method to capture the information about neutron skin thickness as well.
Actually, for a nucleus, the proton and neutron matter distributions are mutually influenced with each other.
Although the neutron skin thickness can be used to ascertain the equation of state (EoS) of isospin asymmetric nuclear matter, the tension between the neutron skin thicknesses of $^{48}$Ca (CREX) and $^{208}$Pb (PREX2) cannot be reconciled in the simulated protocol~\cite{PhysRevLett.129.232501,YUKSEL2023137622,PhysRevC.108.024317,MIYATSU2023138013}.
This means more underlying mechanisms should be captured as much as possible.
Meanwhile, the values of charge radii should also be simultaneously taken into account in evaluating the nuclear EoS through the neutron skin thicknesses of $^{48}$Ca and $^{208}$Pb.

\section{SUMMARY AND OUTLOOK}\label{fourth}
In this work, the influence of neutron-proton correlation deduced from the neutron- and proton-pairs condensation around Fermi surface is firstly incorporated into the Skyrme energy density functionals (EDFs). The modified model can reproduce the trend of changes of differential charge radii in the calcium, nickel, tin, and lead isotopes. Especially, the kink phenomena in charge radii can be significantly reproduced at the corresponding neutron numbers $N=28$, $82$, and $126$ through this modified model. Meanwhile, the shell closure effect at the $N=50$ can also be characterized in the charge radii of nickel and tin isotopes.
Intriguingly, the inverted parabolic-like shape in charge radii of calcium isotopes can be presented between the neutron numbers $N=20$ and $N=28$. This discontinuous evolution can also be predicted in the charge radii between $^{56}$Ni and $^{78}$Ni, but the amplitude is apparently weakened. Although the HFB$^{*}$ model can improve the description of the differential charge radii along Ca, Ni, Sn, and Pb isotopes, the shell closure effects are slightly distorted in the Ca and Pb isotopes.
Besides, in the Pb isotopes, the slightly oscillated $\Delta_{r}^{(2N)}$ values cannot be described well due to the lack of shape deformation toward neutron-deficient regions.
The results obtained by the relativistic EDF models, namely the RHB and RHB$^{*}$ methods, are also shown for the quantitative comparison. The RHB$^{*}$ model can also make an improved description in reproducing the shell closure phenomena of the differential charge radii along Ca, Ni, and Sn isotopic chains, but the shell closure effect at the $N=126$ is excessively emphasized along Pb isotopic chain.

The mean-square charge radius of a nucleus is naturally determined through the charge density distributions.
Recent study demonstrates that the information about charge radii can be derived from the charge-changing cross section measurements of neutron-rich nuclei~\cite{ZHANG20241647,ZHAO2023138269}.
Moreover, the difference of charge-changing cross section of mirror nuclei can also provide an alternative approach to evaluate the equation of state of nuclear matter~\cite{XU2022137333}.
Reliable description of nuclear charge radii provides deep insights into our understanding of fundamental interactions at the extreme neutron-to-proton ratios.
The neutron-proton interactions originating from the valence neutrons and protons can describe the fine structure of nuclear charge radii along a long isotopic chain~\cite{PhysRevLett.58.658,Angeli_1991,Sheng2015,PhysRevC.105.014308,DONG2023137726}.
Moreover, as demonstrated in Refs.~\cite{MILLER2019360,CAURIER2001240,COSYN2021136526}, more extra isospin interactions coming from the neutron-proton correlations should be considered properly in accessing the nuclear size.
This suggests that neutron-proton correlations play an indispensable role in determining the discontinuous behaviors of nuclear charge radii throughout the whole nuclide chart.
As demonstrated in Refs.~\cite{PhysRevC.105.L021301,HU2024138969}, the precise data on mirror charge radii cannot make a rigorous constraint on the slope parameter $L$, even the worse correlation can be obtained between the mirror charge radii difference $\Delta{R}_{\mathrm{ch}}$ and the slope parameter $L$.
It should be noted that the neutron-pairs condensation around Fermi surface has an impact on determining the charge radii.
This means that the influence of neutron- and proton-pairs condensation around Fermi surface should be considered properly in constraining the equation of state of isospin asymmetric nuclear matter.

\section{Acknowledgements}
This work was supported by the Natural Science Foundation of Ningxia Province, China (No. 2024AAC03015), the Central Government Guidance Funds for Local Scientific and Technological Development, China (No. Guike ZY22096024), the Open Project of Guangxi Key Laboratory of Nuclear Physics and Nuclear Technology, No. NLK2023-05, and the Key Laboratory of Beam Technology of Ministry of Education, China (No. BEAM2024G04). X. J. was grateful for the support of the National Natural Science Foundation of China under Grants No. 11705118, No. 12175151, and the Major Project of the GuangDong Basic and Applied Basic Research Foundation (2021B0301030006). N. T. was grateful for the support of the key research and development project of Ningxia (Grants No. 2024BEH04090) and the Key Laboratory of Beam Technology of Ministry of Education, China (No. BEAM2024G05).
L.-G. C. was grateful for the support of the National Natural Science Foundation of China under Grants No. 12275025 and No. 11975096, and the Fundamental Research Funds for the Central Universities (2020NTST06).
F.-S. Z. was supported by the National Key R$\&$D Program of China under Grant No. 2023YFA1606401 and the National Natural Science Foundation of China under Grants No. 12135004.

\bibliography{refsanw}

\end{document}